\documentclass{nature}

\usepackage{color} 
\usepackage{footnote}
\makesavenoteenv{tabular}
\makesavenoteenv{table}
\usepackage{amsfonts}
\usepackage{pdfpages}
\usepackage[]{algorithm2e}
\usepackage{amsmath}
\usepackage[T1]{fontenc}
\usepackage{lmodern}
\usepackage[normalem]{ulem}
\usepackage{color,soul}

\usepackage{graphicx}
\makeatletter
\let\saved@includegraphics\includegraphics
\AtBeginDocument{\let\includegraphics\saved@includegraphics}
\renewenvironment*{figure}{\@float{figure}}{\end@float}
\makeatother

\title{Quasi-spectral characterization of intracellular regions in bright-field light microscopy images}

\author{Kirill Lonhus, Renata Rycht\'{a}rikov\'{a}, Ganna Platonova \& Dalibor \v{S}tys}

\begin{document}
\maketitle

\begin{affiliations}
 \item University of South Bohemia in \v{C}esk\'{e} Bud\v{e}jovice, Faculty of Fisheries and Protection of Waters, South Bohemian Research Center of Aquaculture and Biodiversity of Hydrocenoses, Kompetenzzentrum MechanoBiologie in Regenerativer Medizin, Institute of Complex Systems, Z\'{a}mek 136, 373 33 Nov\'{e} Hrady, Czech Republic
\end{affiliations}

\begin{abstract}
Investigation of cell structure is hardly imaginable without bright-field microscopy. Numerous modifications such as depth-wise scanning or videoenhancement make this method being state-of-the-art. This raises a question what maximal information can be extracted from ordinary (but well acquired) bright-field images in a model-free way. Here we introduce a method of a physically correct extraction of features for each pixel when these features resemble a transparency spectrum. The method is compatible with existent ordinary bright-field microscopes and requires mathematically sophisticated data processing. Unsupervised clustering of the spectra yields reasonable semantic segmentation of unstained living cells without any \textit{a priori} information about their structures. Despite the lack of reference data (to prove strictly that the proposed feature vectors coincide with transparency), we believe that this method is the right approach to an intracellular (semi)quantitative and qualitative chemical analysis.
\end{abstract}

\section*{INTRODUCTION}

Bright-field microscopy in videoenhancement mode shows an unprecedented success as a method of living object investigation since it is cheap and non-intrusive in preparation of samples, and, in its innovative set-up,\cite{Rychtarikova2017} has an excellent spatial and temporal resolution, which opens many possibilities for automation. Classical image-processing techniques such as feature extraction or convolution neural networks do not work so well due to huge variability in microworld data. It calls for image pre-processing techniques that would utilize all available information to supply rich, physically relevant feature vectors in subsequent methods of analysis.

Indeed, classical bright-field microscopy measures properties of incoming light affected by a sample. If multi-photon processes are negligible and, then, intensities are reasonable, a linear response model can be used. Then, a medium observed in such a model can be fully characterized by a transparency spectrum $T(\vec{r})$ defined for each pixel. Such a spectrum can give ultimate information about the medium and boost subsequent machine learning methods significantly.

The most convenient, classical way of obtaining such a spectrum is to modify a measuring device (microscope). It is mostly done using single scanning interferometers,\cite{Lindner2016} matrices of them,\cite{Heist2018} matrices of color filter arrays\cite{Wu2016}, or other adjustable media.\cite{Wachman2014,Dahlberg2016} Such technical arrangements
can be further successfully coupled with machine learning methods as well.\cite{Zhu2018} Purely instrumental methods are certainly the most correct but require sophisticated equipment and are not fully compatible with typical bright-field techniques like depth-wise z-scanning. Due to both hardware and algorithms, this makes these methods rather a separated group than a subtype of the bright-field methods.

For classical bright-field microscopy, the most approaches rely on trained (or fitted) models based on a set of reference images with known properties.\cite{Garini2006} Most mature methods rely on the principal component analysis\cite{Maloney1986} or sparse spatial features.\cite{Parmar2008} Some of such techniques do not aim to full-spectral reconstruction but rather to a more effective colour resolution (which has been very useful in distinguishing fluorescence peaks).\cite{Wang2019} The main disadvantage of such methods is the global approach, which is feasible only for homogeneous images. Most "local" methods include different artificial neural networks\cite{Alvarez-Gila2017} and can work well if they are trained with a reference dataset that is similar to the observed system. The data of this kind almost never occurs in microscopy due to bigger variability of objects in microworld (for the reason that, e.g., known objects are artificial, an investigated system is living, or the in-focus position can be ambiguous). This gives a cutting edge to physically inspired methods which make no assumption about type of observed object and does not use special equipment except of a classical bright-field microscope.

\section*{Theoretical model}

\begin{figure}
\centering
\includegraphics[width=0.7\textwidth]{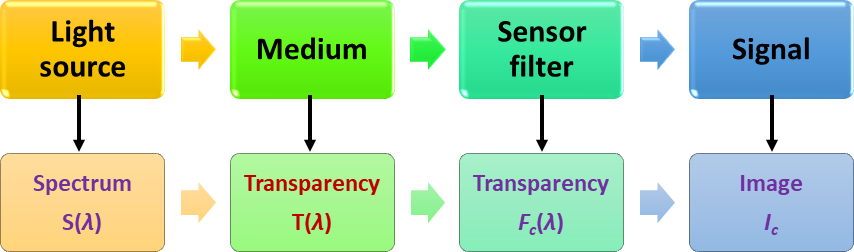}
\caption{Measurement process model.}
\label{Fig1}
\end{figure}

For most biologically relevant objects multi-photon interactions can be neglected.\cite{Hoover2013} Thus, a linear response model can be used for description of the measurement process. The model consists of four entities (Fig.~\ref{Fig1}) which are physically characterized as follows:
\begin{enumerate}
\item \textbf{Light source} gives a light spectrum $S(\lambda)$, which is assumed constant and spatially homogeneous.
\item \textbf{Medium} is, in each point of the projection onto a camera sensor plane, characterized by an unknown transparency spectrum $T(x, y, \lambda)$.
\item \textbf{Camera filter}, where each camera channel $c$ is characterized by a quantum efficiency curve $F_c(\lambda)$.
\item \textbf{Camera sensor} is described (by purely phenomenological approach) by exposure time $t_e$ and energy load curve $I_c = f(E)$, where $I_c$ is the pixel sensor output (intensity) and $E$ is energy absorbed by the pixel sensor during the exposure time. We assume that the image is not saturated and, thus, $f(E)$ can be approximated linearly.
\end{enumerate}
Mathematically, it can be expressed as
\begin{equation}
I_c = f\cdot \int_{0}^{t_e} \int_{\lambda_{min}}^{\lambda_{max}} S(\lambda)\cdot T(\lambda)\cdot F_{c}(\lambda)\cdot d\lambda\cdot dt,
\label{eq:process_compl}
\end{equation}
where $I_c$ is the image intensity at a given pixel. All observable, biologically relevant, processes are slow compared with the camera exposure time (usually in a few ms) and, therefore, the outer integral can be eliminated. More importantly, let variable $f$, which reflects the dependence between the spectral energy and the sensor response, be 1. The multiplication inside the internal integral is commutative, which allows us to introduce an effective incoming light $L_c(\lambda) = S(\lambda)\cdot F_c(\lambda)$. These all mathematical treatments give the reduced equation for the measurement process as
\begin{equation}
\label{eq:process_red}
I_c = \int_{\lambda_{min}}^{\lambda_{max}} L_c(\lambda)\cdot T(\lambda)\cdot d\lambda.
\end{equation}

Intentionally, this simple model does not include any properties of optics, sophisticated models of light-matter interactions, and spatial components (focus, sample surface, etc.). The aim of the method is to describe an observed object in the best way, with minimal assumptions on its nature or features.

\section*{Model extension for continuous media}
In order to extract a transparency profile from the proposed model, one has to solve an inverse problem for a system of 3 integral equations (in case of a 3-channel, RGB, camera). This cannot be solved directly, since the model is heavily underdetermined. (In this text, by terms "transparency" and "spectrum" we mean "quasi-transparency" and "quasi-spectrum" since this method determines only the properties of a microscopy image which are similar to the transparency spectra but not the transparency itself.)

Additional information can be squeezed from the physical meaning of the observed image -- neighbouring pixels are not fully independent. The observed object usually has no purely vertical parts (which is quite typical for cell-like structures) and other Z-axis related changes are not fast.\cite{Lugagne2018} If this holds, the image can be treated as a continuous projection of the object's surface (in optical meaning) onto the camera sensor. In this case, the neighbouring pixels correspond to neighbouring points in the object. 

In addition, let us assume that the object's volume can be divided into subvolumes in a way that the transparency spectra inside a subvolume will be spatially continuous (in L2 meaning). This assumption is quite weak, because it can be satisfied only if the volumetric image has a subvolume of the size which is equal to the voxel size. 

For biological samples which show almost no strong gradients of structural changes holds that the pixel demarcates the projected image. Formally, this criterion can be expressed as
\begin{equation}
\label{eq:cont}
\int_{\lambda_{min}}^{\lambda_{max}}\lvert T(\vec{r}, \lambda) - T(\vec{r} + \vec{u}, \lambda) \rvert^2 d\lambda < q, \;\;\; \forall \lvert \vec{u}\rvert < \epsilon,
\end{equation}  
where $\vec{u}$ is a random vector and $q, \epsilon$ are small numbers. This equation closely resembles the Lyapunov stability criterion. The $\epsilon$ reflects the neighbourhood size and $q$ is related to the degree of discontinuousness. It can be violated, if $\vec{u}$ crosses a border between objects, but not inside a single object.

\section*{Optimization procedure}

For pixel $m$, the combination of optimization criteria in Eqs. 2--3 gives (in discrete form)
\begin{equation}
\label{eq:optima}
F_{m} = \sum_{c=1}^C e^{\lvert \int_{\lambda_{0}}^{\lambda_{w}}L_c(\lambda)\cdot T_{m}(\lambda)d\lambda \; - \; \\I_{m}\rvert} - C + \frac{1}{N}\sum_{n\in \mathbb{N}_m}G_{mn}\sum_{i=1}^{w}[T_m(\lambda_i) - T_n(\lambda_i)]^2,
\end{equation} 
where $C$ is the number of channels, $w$ is the number of discrete wavelengths, $G_{mn}$ is a measure of discontinuousness between pixels $m$ and $n$. The $\mathbb{N}_m$ is a set of points, which have the Euclidean distance to the pixel $m$ equal or less than $\mathcal{T}_{ED}$. Authors used $\mathcal{T}_{ED}=1$, but a larger neighbourhood may improve convergence speed. The integral in the first part of Eq.~4 is supposed to be solved numerically. Authors used the Simpson integration method\cite{Velleman2005} with discretization $\lvert\lvert \lambda_i\rvert\rvert = 48$.

The trickiest issue in Eq. \ref{eq:optima} is calculation of discontinuousness measure $G_{mn}$. We defined it as
\begin{equation}
\label{eq:disc}
G_{mn} = \frac{1}{L_{mn}}\prod_{k\in \mathbb{B}_{mn}} \{[\mathcal{E}_{k} = 0] + [\mathcal{E}_k \neq 0]\cdot (1 - \mathcal{T}_b)\cdot (1 - D_{k})\},
\end{equation} 
where $D_{k}$ is a central gradient in pixel $k$, $\mathcal{T}_b$ is a bias parameter (authors used $\mathcal{T}_b = 0.9$), and $\mathbb{B}_{mn}$ is a set of points, which form lines between pixels $m$ and $n$. The set of such points is calculated using the Bresenham algorithm.\cite{Kuzmin1995} The $\mathcal{E}_k$ indicates whether pixel $k$ is classified as an edge. For this we used the Canny edge detection algorithm\cite{Canny1986} applied to a gradient matrix smoothed by a 2D Gaussian filter with the standard deviation equal\cite{Elboher2012} to 0.5. 

The gradient calculation is different for the first and further iterations. In the first iteration, there is no valid spectral guess, and the gradients and the edge detection are calculated for the original image. The used edge detection algorithm requires a single-channel (grayscale) image, however, the input image is RGB. We used the principal component analysis (PCA)\cite{Pearson1901,SouthAlabamaHair2018} and retained only the first principal component in order to obtain the maximal information on the grayscale representation of data.

In the non-first iterations, there is a spectral guess and, instead of the gradient, we used the cross-correlation with zero lag: $D_{k} = T_{k-1}(\lambda)\star T_{k+1}(\lambda)$. The vertical and horizontal gradient were merged by the Euclidean norm.

\begin{figure}
\centering
\includegraphics[width=0.7\textwidth]{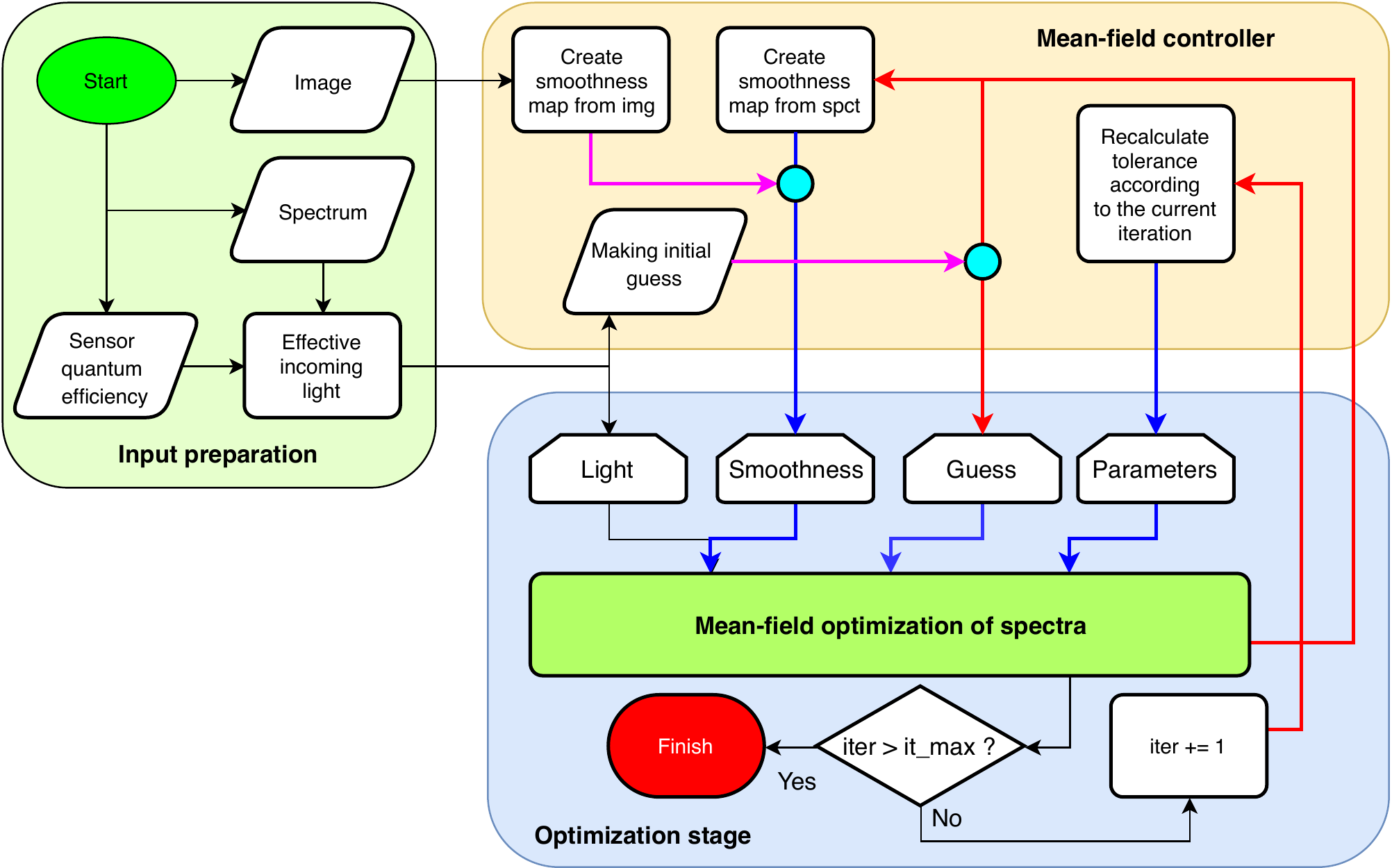}
\caption{The flow chart of the method. The magenta lines denote the routes for the 1$^{\mbox{st}}$ iteration. The red and blue lines show the direct and indirect feedback between iterations, respectively.}
\label{Fig2}
\end{figure}

For numerical optimization of Eq. 4, the covariance matrix adaptation evolution strategy (CMA-ES)\cite{Hansen2001} was proved to be a suitable robust global optimization method.\cite{Hansena} Due to the mean-field nature of the second part of Eq. 4, the method is iterative with, usually, 20--40 iterations to converge. In each iteration step and for each pixel, the minimization is conducted until a predefined value of loss function is achieved. Different schedules of tolerance changes can be applied, authors used the simplest one -- linear decrease. The algorithm flow chart is presented in Fig.~\ref{Fig2}.

\section*{Microscopy system and camera calibration}

In order to obtain reasonable local spectra, we must ensure that camera sensor pixels have homogeneous responses. From hardware point of view, they are printed as semiconductor structures and cannot be changed. Therefore, we introduced a spectral calibration in the form of post-processing routine, which is designed for obtaining equal responses from all camera pixels.

\begin{figure}[htbp]
\centering
\includegraphics[width= .85\textwidth]{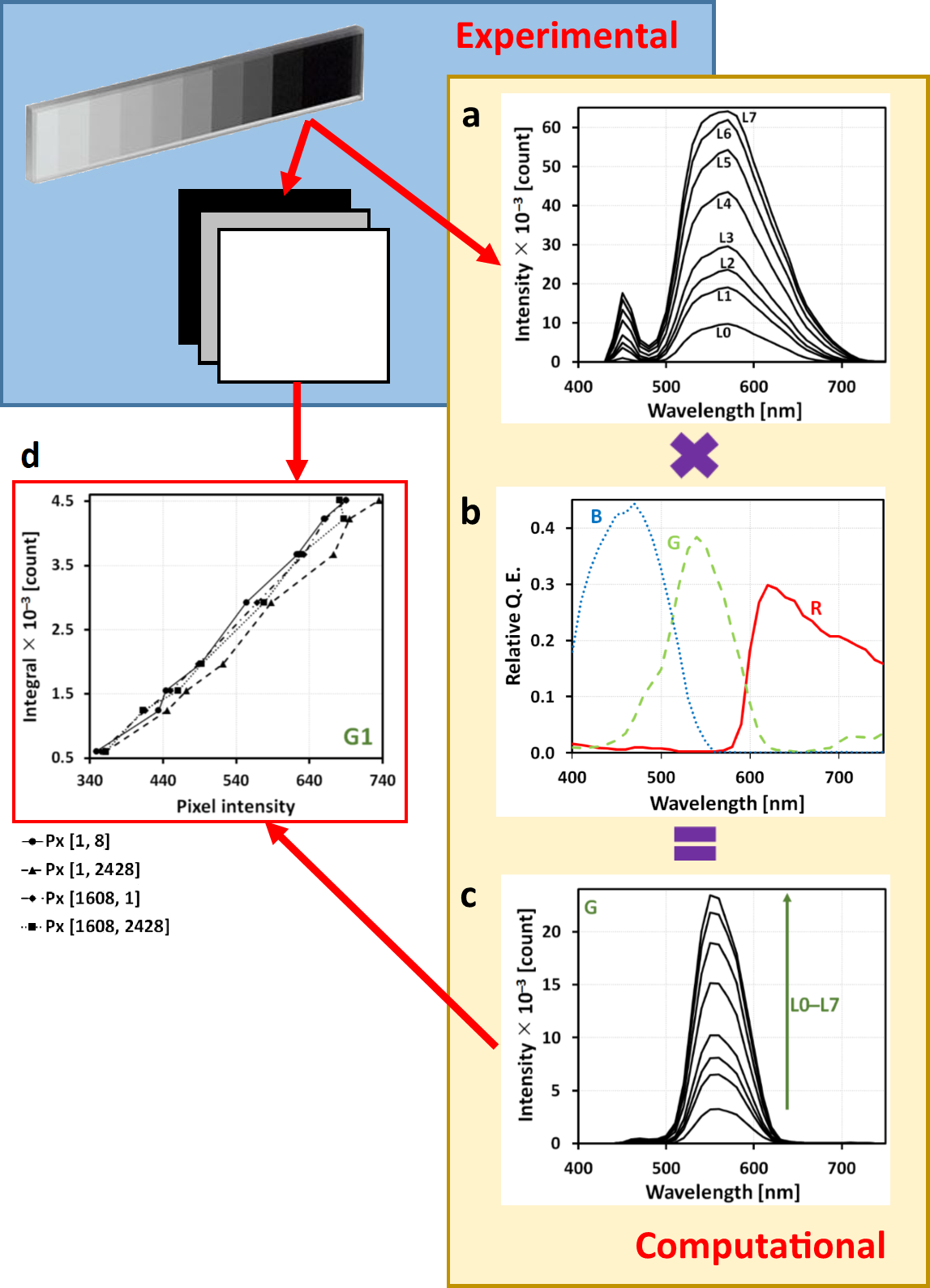}
\caption{(a) Light spectra of grayscale layers measured by a fibre spectrophotometer, (b) declared spectra of RGB camera filters, (c) calculated spectra of incoming light reaching the blue camera channel. The integral under the curve (c) was used as a calibration value for the construction of the calibration curve. (d) Calibration curves for selected blue camera pixels lying in the same column (pixel indices are depicted).}
\label{Fig3}
\end{figure}

The first part of calibration is experimental and aimed at measuring each pixel's sensitivity. We took a photograph of the background through a set of gray layers with varying transparency, covering a 2-mm thick glass (type Step ND Filter NDL–10S–4). After that, we replaced the microscope objective by a fibre of a spectrophotometer (Ocean Optics USB 4000 VIS-NIR-ES) to record spectra corresponding to each of the filters, see Fig.~\ref{Fig3}a.

The second part is computational. For each pixel, we constructed a piece-wise function $S(M)$, where $S$ is an integral of the spectrum measured by the fibre spectrometer in each point $S_i$ and $M_i$ is an intensity value in the image. Between these points, the function $S(M)$ is linearly interpolated, see Fig.~\ref{Fig3}d. For a colour camera that we used, the algorithm is slightly different. Most of the RGB cameras are equipped with a Bayer filter, which effectively discriminates 3 sorts of pixels. Each `sort' has a different dependence of the quantum efficiency on the wavelength, see Fig.~\ref{Fig3}b. These dependencies are usually supplied by the camera producer. In this case, the recorded spectrum should be multiplied by the corresponding efficiency curve \textit{prior to} the integration. The result of the multiplication is shown in Fig.~\ref{Fig3}c.

The proposed method of calibration is universal, applicable to any camera producing raw data, and is not based on any assumption about nature of image or underlying acquisition processes. The algorithm itself is post-processing technique and requires calibration images and data from spectrometer. All results described below were obtained after this image correction. The calibration and correction routines are implemented as a native application and are freely available.

\section*{RESULTS}
The method essentially requires only 3 specific inputs: an image, incoming light spectrum, and camera filter profiles. The camera filter profiles are usually supplied with the camera or can be measured directly using an adjustable monochromatic light source. The incoming light spectrum is less straightforward, because the light emitted by the source is somehow altered by the light path. A convenient way is to replace the objective inlet by a cosine corrector with a spectrometer and measure the incident light spectrum. This implies that, in case of any substantial changes in the optical path (e.g., like the objective replacement), the incoming light spectrum has to be remeasured. In practice, it makes no problem to measure a set of spectra corresponding to a different objective, iris settings, etc. 

The proposed method appears to be quite robust to parametrization inaccuracies and errors. We used the quantum efficiency curves supplied by the vendor and measured the spectrum, which is reaching the sample, and obtained practically feasible results. The method can be applied to any bright-field microscope set-up. The only condition is to access the camera primary signal immediately after the analog-to-digital conversion, before some kind of thresholding, white-balancing, gamma correction, or another visual improvement is employed.

\begin{figure}
\centering
\includegraphics[width=1.0\textwidth]{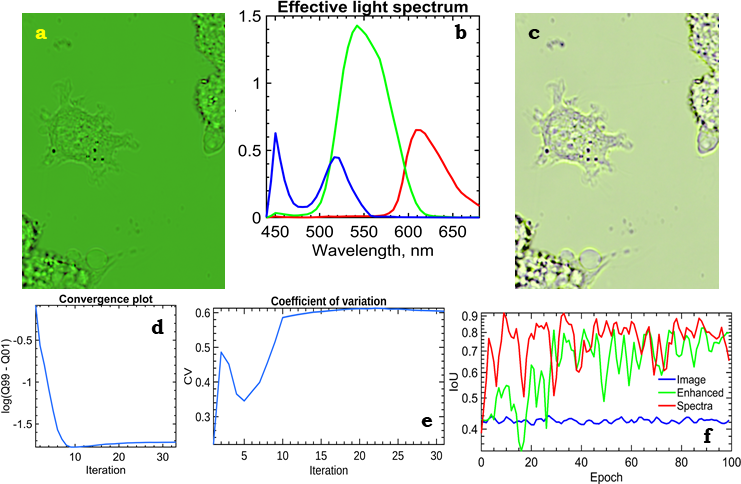}
\caption{The method of quasi-spectra extraction was applied to a raw image of a live cell from a bright-field wide-field light microscope (\textbf{a}) combined with the effective light spectra (\textbf{b}). The cost (\textbf{d}) and variation coefficient (\textbf{e}) demonstrate a quite non-monotonous behaviour. This implies a self-organization of the model. After the reconstruction of the transparency spectra, the image can be viewed under arbitrary illumination such as the absolute black body with T = 5800 K (\textbf{c}). Comparison of the quality of U-Net supervised segmentation for original (raw), contrast-enhanced, and quasi-spectral images (\textbf{f}) shows advantages of the proposed quasi-spectral approach.}
\label{Fig4}
\end{figure}

The sample has to obey 3 assumptions: localized gradients, reasonable flatness, and linear response. If these assumptions hold, the obtained results will be in agreement with physical properties of the medium. Most of relatively flat biological samples (e.g., a single layer of cells) fulfil all these criteria. In order to show the capacity of the method, we used it for analysis of images of unstained live L929 mouse fibroblasts recorded using a video-enhanced bright-field wide-field light microscope in time lapse and with through-focusing. For determination of the best focal position in the z-stack, we used the graylevel local variance.\cite{Pech-Pacheco} The effective light spectrum as the result of multiplication of the light source spectrum by the camera filter transparency curves is shown in Fig.~\ref{Fig4}b. The original raw image is shown in Fig.~\ref{Fig4}a and looks greenish due to the prevalence of green colour in incoming light spectrum.

As clearly seen in Fig.~\ref{Fig4}d--e, the method has a non-trivial convergence behaviour of the variation coefficient (with the local maximum at iteration 2 and the local minimum at iteration 4) and of the cost. The behaviour of the iteration computing process is not related to changes in the schedule of tolerances. This behaviour in iteration process is linearly decreasing until iteration 10, and then is kept constant and the iteration process is stopped if the value of change is 0.01. We have not investigated the reason for this course deeply, but it is definitely repeatable for all the tested measurements (e.g., Fig. S3b--c). A natural way of visual verification of an image of transparency spectra is artificial illumination. We used a spectrum of the black body at T = 5800 K according to the Planck Law (Fig.~\ref{Fig4}c, S3a). The transformed image is quite similar to the raw data, which supports the method validity. To obtain such an image, we multiplied each pixel's transparency spectra by the illumination spectrum and the CIE standard matching curves. The integrals of the corresponding curves gave coordinates in the CIE 1931 colour space.

\begin{figure}
\centering
\includegraphics[width=0.9\textwidth]{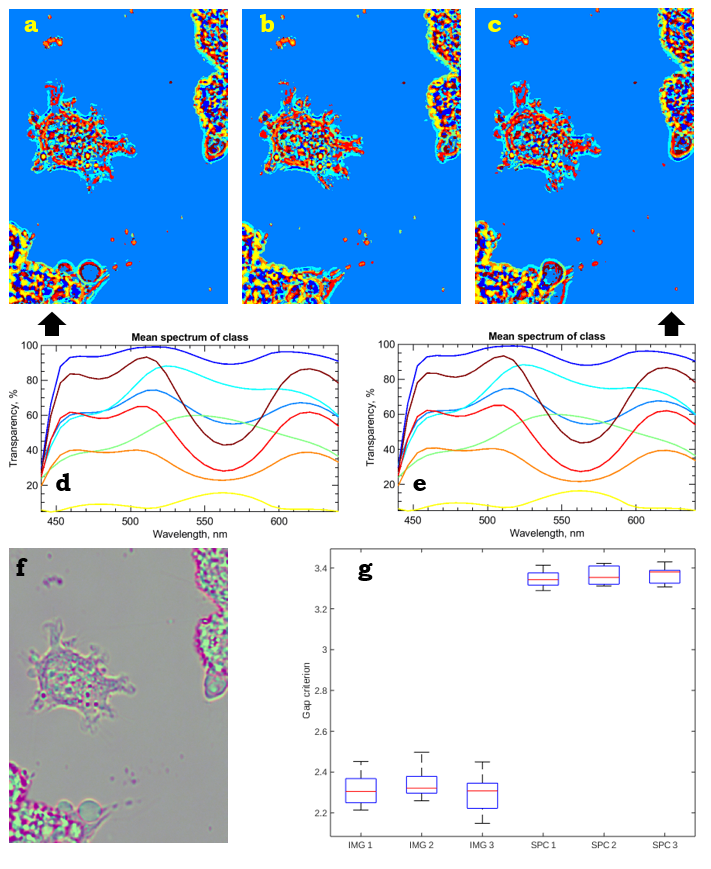}
\caption{A live cell L929 in time lapse (\textbf{a, b, c}) at $k$-means clusterization, $k=10$. The corresponding mean spectra of classes for images (\textbf{a, c}) are shown in (\textbf{d, e}). These spectra are pretty much similar, despite the different images. The gap criteria for the raw data and the relevant spectral counterparts are presented in (\textbf{g}). Dimensionality reduction techniques, e.g., PCA, can be used for better visualization and digital staining (\textbf{f}).}
\label{Fig5}
\end{figure}

Evaluation of the asset of the proposed method of the quasi-spectral reconstrunction (Fig.~\ref{Fig5}a--e) for clusterization against the raw data is quite tricky, because we have no ground truth. But, nevertheless, there are numerous methods of quality estimation for unsupervised learning.\cite{Mekaroonkamon2016} Such methods are usually used for determination of the optimal number of clusters in datasets. Our aim is slightly different --- to compare the accuracy of the clusterization for two datasets with different dimensionality. This naturally yields a choice of metric -- cosine -- since this metric is normalized and not affected by magnitude to such an extent as the Euclidean metric. Another fact that can be utilized from the data is that each single image provides 10$^5$--10$^6$ points. It enables us to use a distribution-based method for estimation of clustering accuracy. One of the most general method from this family is gap statistics,\cite{Tibshirani2001} which is reported to perform well and robust even on noisy data, if a sufficient number of  samples is present.\cite{Mekaroonkamon2016} As the clusterization method itself, we used $k$-means with 10 clusters and the cosine metric. Figure~\ref{Fig5}g shows gap criteria for time-lapse raw images and relevant spectral counterparts. The proposed method leads to better and more stable clustering concurrently. We also investigated different dimensionality reduction techniques (namely PCA,\cite{Pearson1901} Factor Analysis,\cite{STEPHENSON1935} and NNM\cite{Lee1999}), which can be applied before the clustering, but these techniques did not bring any improvement in cluster quality. Despite that, these techniques can be used, e.g., for digital staining and highlighting the details in objects, see Fig.~\ref{Fig5}f. 

In order to verify the benefits of the clusterization of the obtained spectra using $k$-means against the direct image clusterization, simple phantom experiments on microphotographs of oil-air and egg protein-air interface, respectively, were conducted. These phantom experiments showed that the spectral clusterization resulted in both a higher cluster accuracy and a lower variation. Moreover, in order to prove the capacity of the method, we applied supervised segmentation, namely a classical semantic segmentation network, U-Net.\cite{Ronneberger2015} It is a symmetric encoder-decoder convolution network with skip connections, designed for pixel-wise segmentation of medical data. One of the strongest advantage of this network is a very low amount of data needed for successful learning (only a few images can be sufficient for this purpose). We employed 6 images for the network training and 1 image for the method validation. To avoid the data overfitting in the training phase, aggressive dropout (0.5, after each convolution layer) and intensive image augmentation (in detail in Suppl. Material 1) was rendered. We compared the performance of the U-Net network for the original raw images, contrast-enhanced images, and spectral images (Fig.~\ref{Fig4}f). The results of segmentation for the spectral images showed a significantly ($>10\%$) increased accuracy, intersection over union (IoU) 0.9, and a faster convergence speed (8 epochs vs. 40 epochs for contrast-enhanced images). The results were stable to changes in the training and test sets (even when using a single validation image or a set of augmented images derived from validation as mentioned above).

\section*{DISCUSSION}

The primary aim of the method is, in the best possible way, to characterize individual cell parts physically (by a colour spectrum) and, consequently, identify them as different cell regions. Currently, the standard approach for the recognition of organelles is fluorescent (or other dye) staining. In unstained cells, identity of an organelle is guessed from its shape and position. Our approach gives the promise to be able to identify the organelles according to their spectra. However, in order to obtain the same spectra for cells of different samples, full reproducibility of the whole experiment such as optical properties of a Petri dish, thickness and colour of cultivation medium has to be ensured.

An important issue that we have not investigated yet is the influence of sample thickness. The question remains what is the identity of the spectrum if the sample has a non-zero thickness. In Rycht\'{a}rikov\'{a} et al. (2017),\cite{Rychtarikova2017} we showed that the position of the effective focus differs even with the usage of a fully apochromatic lens. This is the biggest complication in interpreting the spectrum. In case of a relatively thick and homogeneous organelle it can be assumed that, in the centre of the focus, the contribution from geometrically different levels are similar. The full answer to this question would be given by a complete 3D analysis that has to be theoretically based on completely new algorithms and is currently out of the possibilities of our computing capacity. To this point, however, we allow to claim that the thickness of the sample affects mainly the integrals below the spectra, not the shapes of the spectra themselves. The usage of the cosine metric, which is, in effect, the angle between distance vectors and is insensitive to the magnitude, would help to mitigate this problem.


It is worth mentioning that, for some real-life biological samples, the measurement model can be violated. We implicitly assume that light intensity reaching the camera chip is always lower than at the time of its production by a light source. The transparency coefficient is bounded by the range $[0, 1]$. Indeed, this is not always true because the sample can contain light-condensing objects (most of these objects are bubbles or vacuoles) which act as micro-lenses. It does not break the method generally but, due to inability to fulfil Eq.~\ref{eq:process_red}, the local optimization gives an abnormally high cost. Such objects should be eliminated from a subsequent analysis because their quasi-spectra are unreliable. After excluding those dubious regions (which occupy only a very small part of the image, provided they are present at all), the rest of the image can be analysed in an ordinary way.

The obtained quasi-spectra should not be considered as object features but are rather imaging process features. Due to the model-free nature of the method, the obtained classes reflect the observed data, not the internal structures of the objects. We think that the convenient bridge between the observed, phenomenological, spectra and the structure is machine learning, since it shows advantage of enormously good statistics (10$^5$--10$^6$ samples per image) and compensate influence of the complicated shape.

\section*{CONCLUSIONS}
This novel method of extraction of quasi-spectra aims at a very challenging problem, which cannot be solved precisely even in theory: some information is irrecoverably lost. The method arises from very general assumptions on the measurement system. The method does not rely on any light-media interaction model or physical properties of the system, which makes this method quite universal. The obtained spectra are applicable in practice for visualization and automatic segmentation task. We intentionally did not consider questions of voxel spectrum, Z-stack spectral behaviour, and meaning of the compromised focus in order to keep the method and its application simple. We pose the described method as an ultimate information squeezing tool, which is a nearly model-free way how to compress the colour and spatial information into representation of the physically relevant features. We believe that, in the future, the method will find its use in robust, mainly qualitative, (bio)chemical analysis.

\section*{MICROSCOPY DATA ACQUISITION}
\subsection{\textbf{Sample preparation}} \mbox{}\\
A L929 (mouse fibroblast, Sigma-Aldrich, cat. No. 85011425) cell line was grown at low optical density overnight at 37$^\circ$C, 5\% $\mbox{CO}_2$, and 90\% RH. The nutrient solution consisted of DMEM (87.7\%) with high glucose (>1 g L$^{-1}$), fetal bovine serum (10\%),  antibiotics and antimycotics (1\%), L-glutamine (1\%), and gentamicin (0.3\%; all purchased from Biowest, Nuaill\'{e}, France).

Cells fixation was conducted in a tissue dish. The nutrient medium was sucked out and the cells were rinsed by PBS. Then, the cells were treated by glutaraldehyde (3\%) for 5 min in order to fix cells in a gentle mode (without any substantial modifications in cell morphology) followed by washing in phosphate buffer (0.2 mol L$^{-1}$, pH 7.2) two times, always for 5 min. The cell fixation was finished by dewatering the sample in a concentration gradient of ethanol (50\%, 60\%, and 70\%) when each concentration was in contact with the sample for 5 min.

The time-lapse part of the experiment was conducted with living cells of the same type. 

\subsection{\textbf{Bright-field wide-field videoenhanced microscopy}} \mbox{}\\
The cells were captured using a custom-made inverted high-resolved bright-field wide-field light microscope enabling observation of sub-microscopic objects (ICS FFPW, Nov\'{e} Hrady, Czech Republic).\cite{Rychtarikova2017} The optical path starts by two Luminus CSM-360 light emitting diodes charged by the current up to 5000 mA (in the described experiments, the current was 4500 mA; according to the LED producer, the forward voltage was 13.25 V which gave the power of 59.625 W) which illuminate the sample by series of light flashes (with the mode of light 0.2261 s--dark 0.0969 s) in a gentle mode and enable the videoenhancement.\cite{Irene} The microscope optical system was further facilitated by infrared 775 nm short-pass and ultraviolet 450 nm long-pass filters (Edmund Optics). After passing through a sample, light reached an objective Nikon (in case of the live cells, CFI Plan Achromat 40$\times$, N.A. 0.65, W.D. 0.56 mm; in case of the fixed cells, LWD 40$\times$, Ph1 ADL, $\infty$/1.2, N.A. 0.55, W.D. 2.1 mm). A Mitutoyo tubus lens (5$\times$) and a projective lens (2$\times$) magnify and project the image on a JAI camera with a 12-bpc colour Kodak KAI-16000 digital camera chip of 4872$\times$3248 resolution (camera gain 0, offset 300, and exposure 293.6 ms). At this total magnification, the size of the object projected on the camera pixel is 36 nm. The process of capturing the primary signal was controlled by a custom-made control software. The z-scan was performed automatically by a programmable mechanics with the step size of 100 nm.

\subsection{\textbf{Microscopy image data correction and visualization}} \mbox{}\\
The acquired image data were corrected by simultaneous calibration of the microscope optical path and camera chip as described in Suppl. Material 1. In this way, we obtained the most informative images on spectral properties of the observed cells.

For visualization, very bright pixels which correspond to light-focusing structures in the sample (mostly bubbles that act as micro-lenses) and violate the assumptions of the model of the proposed quasi-spectral method were detected (as 99\% percentile of intensities) and treated as saturated. After their elimination, the rest of intensities was rescaled to the original range.

\section*{LIST OF SYMBOLS}
\begin{tabular}{@{}ll}
$\mathbb{B}_{mn}$ & set of pixels that form lines between pixels $m$ and $n$\\
$c$ & colour of a camera filter or an image channel; for colour camera $c = \{red, green, blue\}$\\
$C$ & number of image channels\\
$D_k$ & central intensity gradient in pixel $k \in \mathbb{B}_{mn}$ in calculation of $G_{mn}$\\
$E$ & energy absorbed by a camera sensor during an exposure time $t_e$\\
$\mathcal{E}_k$ & parameter in computation of $G_{mn}$ which indicates if the pixel $k$ is classified as\\
		& an region edge\\
$f$ & variable which reflects a dependence between the spectral energy and the sensor\\
	& response; $f = 1$\\
$F_c(\lambda)$ & spectral quantum efficiency of a camera filter $c$\\
$F_m$ & spectral quantum efficiency of a pixel $m$\\
$G_{mn}$ & measure of discontinuousness between pixels $m$ and $n$\\
$i$ & label of a discrete wavelength; $i = \{1,2,..., w\}$\\
$\mbox{iter}$ & iteration\\
$\mbox{it\_max}$ & maximal iteration (predetermined)\\
$I_c$ & pixel intensity at colour channel $c$\\
$k$ & pixel in the set $\mathbb{B}_{mn}$\\
$L_c$ & light effectively incoming onto a camera sensor, i.e. onto a camera filter\\
$m, n$ & pixel labels\\
$M_i$ & intensity value in the image\\
$N$ & number of pixels in the set $\mathbb{N}_m$\\
$\mathbb{N}_m$ & set of pixels with the Euclidean distance to the pixel $m$ equal or less than $\mathcal{T}_{ED}$\\
$q$ & parameter related to the degree of discontinuousness in spectral regions\\
$\vec{r}$ & position vector for a pixel at coordinates $(x, y)$\\
$S$ & integral of the spectrum measured by the fibre spectrophotometer in each point $S_i$\\
$S(\lambda)$ & light spectrum of a light source\\
$t_e$ & camera exposure time\\
T & thermodynamic temperature; kelvin [K]\\
$T_m(\lambda_i)$ & transparency spectrum of pixel $m$ at wavelength $\lambda_i$\\
$T_n(\lambda_i)$ & transparency spectrum of pixel $n$ at wavelength $\lambda_i$\\
$T(x, y, \lambda)$ & transparency spectrum of a medium at each pixel in general\\
$\mathcal{T}_b$ & bias parameter in computation of $G_{mn}$; $\mathcal{T}_b = 0.9$\\
$\mathcal{T}_{ED}$ & threshold for the selection of the neighbourhood of pixel $m$, i.e., the Euclidean\\
	& distance between pixels $m$ and $n$; $\mathcal{T}_{ED} = 1$\\
$\vec{u}$ & change of a pixel position vector\\
$w$ & number of discrete wavelengths\\
$x,y$ & vertical and horizontal pixel coordinates\\
$\epsilon$ & parameter which reflects the studied pixel's neighbourhood size in general\\
$\lambda$ & light wavelength; nanometer [nm]\\
\end{tabular}


\section*{REFERENCES}
\bibliographystyle{naturemag}
\bibliography{Lonhusetal}


\begin{addendum}
\item[ACKNOWLEDGEMENTS] This work was supported by the Ministry of Education, Youth and Sports of the Czech Republic---project CENAKVA (LM2018099)---and from the European Regional Development Fund in frame of the project Kompetenzzentrum MechanoBiologie (ATCZ133) and Image Headstart (ATCZ215) in the Interreg V-A Austria--Czech Republic programme. The work was further financed by the GAJU 013/2019/Z project.

\item[AUTHOR CONTRIBUTION]
K.L. is the main author of the paper and of the novel algorithm, G.P. and R.R. are responsible for sample preparations, microscopy data acquisition, and image calibration, R.R. contributed to the text of papers substantially, D.\v{S}. is an inventor of the videoenhanced bright-field wide-field microscope. D.\v{S}. and R.R. lead the research. All authors read the paper and approved its final version.

\item[COMPETING FINANCIAL INTERESTS]

The authors declare that they have no competing financial interests.

\item[Correspondence]

Correspondence and requests for materials should be addressed to K.L.\linebreak (lonhus@frov.jcu.cz).

\item[Supplemental Data]

\item[Data Availability Statement]
The software for quasi-spectral characterization of images, the relevant Matlab codes, the software for image calibration, the U-Net segmentation package, and testing images are available in the supplementary materials at the Dryad Data Depository.\cite{dryad}
\end{addendum}

\includepdf[page=-]{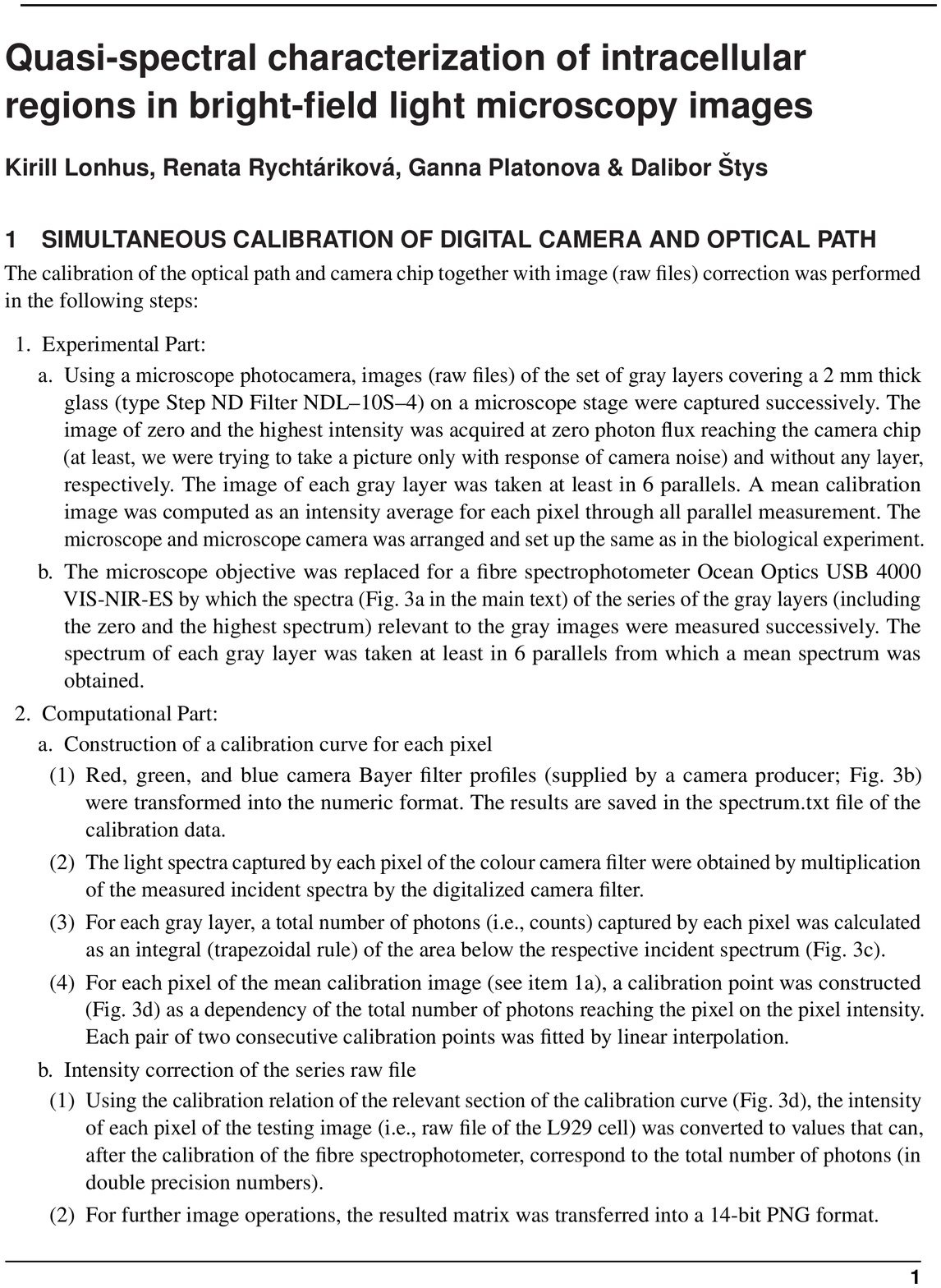}


\end{document}